# Magnetic domain pattern in hierarchically twinned epitaxial Ni-Mn-Ga films

Anett Diestel[a,b,*], Volker Neu[a], Anja Backen[a,b], Ludwig Schultz[a,b] and Sebastian Fähler[a]

[a] IFW Dresden, Institute of Metallic Materials, P.O. Box 270116, 01171 Dresden, Germany

[b] Dresden University of Technology, Institute of Materials Science, 01062 Dresden, Germany

[*] Corresponding author: a.diestel@ifw-dresden.de, Tel.: +49 (0)351 4659-259

*Magnetic shape memory alloys exhibit a hierarchically twinned microstructure, which has been well examined in epitaxial Ni-Mn-Ga films. Here we analyze consequences of this "twin within twins" microstructure on the magnetic domain pattern. Atomic and magnetic force microscopy are used to probe the correlation between the martensitic microstructure and magnetic domains. We examine consequences of different twin boundary orientations with respect to the substrate normal as well as variant boundaries between differently aligned twinned laminates. A detailed micromagnetic analysis is given which describes the influence of the finite film thickness on the formation of magnetic band domains in these multiferroic materials.*

## 1 Introduction

Magnetic Shape Memory (MSM) alloys are multiferroics which exhibit ferroelastic and ferromagnetic order. The coupling between martensitic microstructure and magnetic domains gives rise to the Magnetically Induced Reorientation (MIR) of martensitic variants[1], resulting in huge strains up to 10 % in moderate magnetic fields[2]. The martensitic and the magnetic microstructures have been examined in detail in bulk single crystals. The martensitic microstructure of modulated martensite (e.g. 10M and 14M) exhibits a characteristic "twin within twins" microstructure, which involves several generations and length scales.[3,4] According to *Roytburd* [5], each generation of twins forms in order to compensate the type of elastic energy, which was not minimized by the finer twinning generation. The first generation in this hierarchy can be identified in the modulated unit cell itself.[4] According to the adaptive concept of *Khachaturyan* et al.[6] the modulated unit cell is a nano-twinned microstructure of a simple tetragonal unit cell. The next generation of twinning occurs between differently aligned unit cells of the modulated structure. These twin boundaries are often called mesoscopic twin boundaries.[4] As the MIR effect is based on the movement of this mesoscopic twins by a magnetic field, we will refer to them as twin boundaries throughout this paper.

A martensitic microstructure commonly contains regions of parallel twin boundaries, so-called laminates. There are different possibilities to orient laminates, which are crystallographically equivalent. Thus, a further generation of macroscopic twin boundaries forms, which connects mesoscopic laminates of different orientations. To avoid confusion with the other twin generations, we will call the twin boundaries between different oriented laminates variant boundaries.

Each generation of twin boundaries differs by several orders of magnitude in its twinning period, twin boundary energy[7] and mobility[4]. Accordingly each generation should have a different impact on the magnetic domain pattern. For the first generation, the twinning period is just a few unit cells, which is well below the magnetic exchange length[8]. Accordingly, magnetization is completely coupled between nanotwins. Indeed one can calculate both the sign and absolute value of magnetocrystalline anisotropy of the modulated cell simply as a mean average of the nanotwinned tetragonal building blocks. This approach also gives the appropriate temperature dependency of magnetocrystalline anisotropy.[9]

Consequences of the mesoscopic twinning on the magnetic domain patterns have been extensively examined in bulk single crystals, as the MIR effect is based on this coupling. The fundamental pattern is a staircase domain pattern.[10-12] This pattern is possible due to different symmetries of the underlying ferroelastic or ferromagnetic transformation, resulting in martensitic variants or magnetic domains. While for a martensitic variant only the direction of the magnetic easy *axis* is important, for a magnetic domain also the *direction* of magnetization matters. Thus one can distinguish two different domain walls: The 90°-domain walls coinciding with twin boundaries and are formed, since they allow magnetization to follow the magnetic easy axis to minimize magnetocrystalline anisotropy energy. The 180°-domain walls can be formed within one martensitic variant since there are two possible magnetization directions following the given crystallographic easy axis. This additional freedom for the formation of a magnetic domain pattern is commonly used to minimize magnetostatic energy.

In addition to the fundamental staircase pattern, magnetic domain mirroring at twin boundaries[13] and the formation of spike domains at 90°-domain walls to reduce the magnetic stray field at the surface[14] were observed in single crystalline Ni-Mn-Ga. To our knowledge, there is no report on the domain pattern occurring at variant boundaries.

Another rarely analyzed aspect is the influence of the reduced symmetry of thin films compared to bulk. Up to now magnetic domain patterns of thin films have mostly been examined in polycrystalline form.[15] An increased magnetic domain period with increasing film thickness according to *Kittel*[16] was observed, but the polycrystalline nature hinders probing a possible correlation between magnetic and martensitic microstructure. In a previous work[17], we analyzed the magnetic domain configuration in a thickness series of epitaxial 14M martensitic Ni-Mn-Ga films. As these findings represent one particular case of twin boundary alignment out of six possible orientations, another case of twin boundary alignment is introduced in this paper. In addition, a detailed micromagnetic analysis of the type X domain pattern is given. We also analyze the variant boundaries occurring between both different orientations.

In this paper we examine Ni-Mn-Ga films on a rigid substrate only. The rigid substrate inhibits any macroscopic length change and therefore no substantial changes of the martensitic microstructure are possible. Furthermore, in the present films the martensitic transition temperature is above the Curie temperature. Therefore one may consider the martensitic microstructure as given and can reduce the investigation to the question of how the magnetic domain pattern adapts to it. This is of advantage as the martensitic microstructure of epitaxial films has been examined in detail.[18-21] All measurements have been performed in the as-deposited state at room temperature, which is below the ferroelastic and ferromagnetic order temperature. We did not apply an external magnetic field during the measurements. Hence, these measurements represent the starting point for a future field dependent reorientation.

## 2    Experimental

Ni-Mn-Ga films (sample X and Y) of different film thicknesses ($d_X = 2$ µm; $d_Y = 1.5$ µm) were prepared by DC magnetron sputter deposition from an alloyed target on a heated single crystalline (100)-MgO substrate ($T_X = 400$ °C; $T_Y = 300$ °C), as described in detail in our previous work[22]. The epitaxial relation MgO(100)[001]∥Cr(100)[011]∥Ni-Mn-Ga(100)[011] describes the film architecture. All micrographs shown here are aligned with the MgO[001] substrate edges parallel to the picture edges and the austenitic Ni-Mn-Ga unit cell is rotated by 45°. The sacrificial chromium layer ($d_{Cr} = 100$ nm) improves the film quality and enables the preparation of freestanding Ni-Mn-Ga films.[23] The film compositions (X: $Ni_{47.8}Mn_{32.5}Ga_{19.7}$ and Y: $Ni_{48.5}Mn_{32.8}Ga_{18.8}$) were determined by energy dispersive X-ray spectroscopy (EDX) with an accuracy of 0.5 at.% using a $Ni_{50}Mn_{25}Ga_{25}$ standard. All investigated samples exhibit the 14M modulated martensitic structure, which has been examined by X-ray diffraction and transmission electron microscopy measurements, which will be published elsewhere.

Martensitic and magnetic microstructure was probed by atomic (AFM) and magnetic force microscopy (MFM) using a digital instrument dimension 3100. Topography was imaged by height contrast in tapping mode and magnetic micrographs were scanned in lift mode by a standard magnetic tip with a Co-alloy coating and the magnetization along the tip axis. The lift scan height ranges from 50 to 100 nm depending on the strength of the magnetic stray field.

## 3    Magnetic properties of Ni-Mn-Ga

Due to the technological relevance of magnetic shape memory alloys, the magnetic properties of Ni-Mn-Ga are well examined. For the micromagnetic calculations, we use the following values of 14M martensite at room temperature:[24,25] magnetostatic energy density $K_d = \frac{J_s^2}{2\mu_0} = 0.14$ MJm$^{-3}$ with saturation magnetization $J_S = 0.6$ T, uniaxial anisotropy coefficient $K_u = 0.9 \cdot 10^5$ Jm$^{-3}$. The spin wave stiffness constant $W = 100$ meVÅ$^2$ allows calculating the exchange constant $A = \frac{W \cdot S \cdot N}{2a^3} \approx 6.1$ pJm$^{-1}$ using the magnitude of atomic moment $S = 3.8$ in units of $\mu_B$, the number $N$ of Mn atoms per unit cell and the corresponding lattice parameter $a$ of austenitic Ni-Mn-Ga. From these values the domain wall energy desity $\gamma_{180°} = 4\sqrt{A \cdot K_u} = 2.9$ mJm$^{-2}$ of 180°-domain walls and $\gamma_{90°} \approx 0.3 \cdot \gamma_{180°} = 0.9$ mJm$^{-2}$ of 90°-domain walls are obtained. The latter is calculated through integration of anisotropy and exchange energy density along the profile of a domain wall across a boundary with 90° easy axis orientation (see Eq. 1 in Ref. 26).

## 4 Results and discussion

### 4.1 Orientations of twin boundaries in epitaxial films: type X and Y

Twin boundaries (TB) are well defined crystallographic planes which connect differently oriented martensitic variants. Six different orientations of {101}-type twin boundaries are possible which are sketched in Figure 1 with respect to the cubic austenite unit cell. Each of these orientations represents mesoscopic $a$-$c$-twin boundaries connecting 14M martensitic variants with alternating $a$- and $c$-axis in plane. While in bulk all six possible orientations are equivalent, this is not the case for thin films. Due to the finite film thickness one has to distinguish between two kinds of twinning, which we call type X and Y.

Type X twinned martensite exhibits twin planes inclined by 45° with respect to the substrate normal and their traces run 45° rotated to the substrate edges at the film surface (see Figure 1, orange TB). The crystallographic short $c$-axis lies alternating in- and out-of-plane. Type Y twinned martensite shows twin planes which are aligned perpendicular to the substrate surface with traces running parallel to the substrate edges at the film surfaces (see Figure 1, blue TB). The $c$-axis is aligned in-plane, but alternates between both equivalent orientations 45° rotated to the substrate edges. Type X as well as type Y twinned regions consist of the same kind of modulated 14M martensite, however the $b$-axis and the $a$-$c$-twin boundaries run differently with respect to the sample normal. Since the crystallographic $c$-axis coincides with the magnetic easy axis in 14M martensite[8], substantially different domain pattern in both types of twinning are expected. Only for type X twinning the magnetic easy axis points out-of-plane and stray field effects play an important role. This should be negligible for type Y twinned variants with in-plane magnetic easy axis.

To understand the measurements below it is important to consider that the six different orientations of twin boundaries sketched in Figure 1 are a simplification. As these twin variants form at the irrational habit plane, connecting austenite and martensite, they are tilted and rotated by a few degrees away from precise {101}-planes of austenite.[27] Instead of the six fundamental orientations, a multiple of 24 so-called habit plane variants exist. The slight tilt and rotation is elastically incompatible with the rigid substrate. This incompatibility is compensated by alternating twin variants with positive and negative angular deviations, resulting in a wavy or rhombus-like topographic pattern on the film surface.

In addition to *a-c*-twin boundaries also *a-b*- and *b-c*-twin boundaries are possible due to the orthorhombic distortion of 14M martensite. First indications of these boundaries have recently been reported also for 10M martensite in Ni-Mn-Ga.[28] For the present films we have no indication of these types of boundaries, hence they will not be considered in the following.

### 4.2  Magnetic domain pattern within type X twinning

The correlation between magnetic and martensitic domain structures of type X twinned martensite is explained in detail in our previous work[17]. The key points of this analysis are shortly summarized since they are the basics of the following sections.

The martensitic microstructure of a 2 µm thick Ni-Mn-Ga film (sample X) has been mapped by AFM and is depicted in Figure 2a. A periodical, wavy surface topography with a rhombus-like superstructure is visible. Since in type X the mesoscopic *a-c*-twin boundaries are inclined by 45° towards the substrate normal, the magnetic easy *c*-axis lies alternately in- and out-of-plane (Figure 1, orange TB). The schematical cross sections in Figure 6 illustrate the typical surface profiles of a *c-a*-twinned martensite[22]. Twinning results in a slight inclination between both variants connected by a twin boundary. Traces of twin boundaries are therefore visible on the surface topography as linear ridges and valleys. Due to the epitaxial film growth, the orientation of the austenitic Ni-Mn-Ga unit cell is rotated about 45° in respect to the MgO-substrate edges.[22]

The experimentally observed corresponding magnetic domain pattern (see Figure 2b) consists of magnetic band domains with domain walls (DW) perpendicular to the twin boundaries.[17] For the present 2 µm thick film an additional contrast within the band domains is visible, which resembles the same direction and period of the twinning. This aspect is discussed in detail in section 4.3.

In Figure 2c the correlation between martensitic and magnetic information is summarized by a schematical top view and cross section along MgO[$0\bar{1}1$]. The magnetic easy *c*-axes (black arrows) of neighboring variants lie alternately in- and out-of-plane. Due to the high magnetocrystalline anisotropy of Ni-Mn-Ga the magnetization $\vec{m}$ (orange arrows) follows *c*.

For a given twin boundary orientation (e.g. Figure 2c, cross section: TBs are inclined to the left) there are only two arrangements of magnetization directions possible. The first one is illustrated by orange arrows, where the magnetization alternates between up and right. In the second one the directions of magnetization are reversed to down and left. Any other configurations of magnetization directions (combining up and down or left and right) results in energetically unfavorably charged 90°-DWs.

Since both possible arrangements exhibit a perpendicular magnetization component, which points either up or down, a band domain pattern will form to minimize the magnetic stray field energy.

The same perpendicular alignment of twin boundaries and band domain walls was observed in a thickness series[17] ranging from 125 to 2000 nm. The observed domain period $\Lambda_{DW}$ as a function of the film thickness $d$ follows a square root dependency, as expected from Kittels law[16]: $\Lambda_{DW}(nm) = 25.8 \text{ nm}^{1/2} \cdot d^{1/2}(nm^{1/2})$. Kittels law also predicts appropriate absolute values of $\Lambda_{DW}$, when assuming that within each band domain magnetization is averaged between neighboring twin variants. The validity of this assumption is discussed in detail in section 4.3. This agreement allows considering the observed band domain pattern as the optimum balance between the total domain wall energy and stray field energy. In contrast to these magnetic energies, the twinning period is an optimum balance between elastic energy contributions: total twin boundary energy and elastic energy.[7] Since the equilibrium magnetic domain period exceeds the equilibrium martensitic twinning period, the formation of magnetic band domains parallel to the martensitic variants does not allow obtaining their optimum width. For a perpendicular alignment the domain period does not need to match the twin period and thus a domain period with minimum energy can form.[17] This orthogonal arrangement allows an independent minimization of magnetic and elastic energies, respectively.

A completely different domain pattern is known for bulk samples (staircase domain pattern). The absence of additional 180°-domain walls within the films are due to the unfavorable total magnetic energy. In the following section the observed domain pattern in type X films was analyzed in detail in comparison with competing models.

## 4.3 Micromagnetic analysis of type X domains

In order to explain the observed magnetic domain pattern more precisely, we will compare the micromagnetism of three different patterns, which are sketched in Figure 3: a) a magnetic domain pattern, which considers a homogeneous and tilted magnetization, irrespective of the local anisotropy axis in the underlying martensitic variants, b) a refined model, where magnetization follows the magnetic easy axis within each martensitic variant, and c) a staircase pattern adapted from bulk.

By deriving reasonable approximations for the total magnetic energy of these domain scenarios, we can predict the energetically favored domain state. For discussing magnetostatic energies, the Faraday picture of magnetization is applied. This means that discontinuities in the normalized magnetization vector $\vec{m} = \vec{J}/J_s$ are sources of magnetic charges. There are either surface charges $\sigma = \vec{m} \cdot \vec{n}$ ($\vec{n}$ as normal vector) or volume charges $\rho = -\text{div}(\vec{m})$, which by themselves are sources of stray fields.

In case (a) the considered domain pattern consists of alternating bands with homogeneously tilted magnetization in each domain, which results from averaging over the easy axis of two neighboring martensitic variants (Figure 3a). The total energy consists of the wall energy of 180°-DWs ($\varepsilon_{180°}$), the magnetostatic energy ($\varepsilon_d$) of the tilted parallel magnetic bands with the given domain period ($\Lambda_{DW}$) and the anisotropy energy ($\varepsilon_{ani}$) for rotating the magnetization by 45° out of the local magnetic easy axis throughout the whole film thickness $d$. The respective contributions per unit film area are:

$$\varepsilon_{180°} = \#_{DW} \cdot A_{DW} \cdot \gamma_{180°} = \frac{2}{\Lambda_{DW}} \cdot d \cdot \gamma_{180°} \tag{1a}$$

with $\#_{DW}$ as number and $A_{DW}$ as area of 180°-DWs per unit film surface. According to Ref. 29 the magnetostatic energy of band domains is given by

$$\varepsilon_d = \frac{1.705}{4\pi} \cdot K_d^* \cdot \Lambda_{DW} = \frac{0.8525}{4\pi} \cdot K_d \cdot \Lambda_{DW} \tag{1b}$$

with $K_d^* = (0.5\mu_0) \cdot (J_s^\perp)^2 = 0.5 K_d$, due to the 45° tilt of the magnetization. The anisotropy energy does not depend on the domain wall period:

$$\varepsilon_{ani} = 0.5 \cdot K_u \cdot d \tag{1c}$$

Thus $\Lambda_{DW}$ will be adjusted to minimize the sum of the first two energy contributions only, which leads to the known equation for band domains[16,29]:

$$\Lambda_{DW}^{BD} = \sqrt{\frac{8\pi \cdot \gamma_{180°} \cdot d}{0.8525 \cdot K_d}} \tag{1d}$$

(Note, that Eq. 1d differs by the factor 1.0517 from the equation used in Ref. 17. The factor arises from higher order harmonic terms in the calculation of the magnetostatic energy.) As $\varepsilon_{ani}$ increases with the film thickness $d$, this contribution makes this model more unrealistic for thicker films. The rotation of magnetic moments becomes undesirable and the more complex domain pattern sketched in Figure 3b has to be considered.

In case (b), the magnetization follows the easy axis within each martensitic variant (Figure 3b). The energy contributions of this model are:

$$\varepsilon_{180°} = \#_{DW} \cdot A_{DW} \cdot \gamma_{180°} = \frac{2}{\Lambda_{DW}} \cdot d \cdot \gamma_{180°} \tag{2a}$$

$$\varepsilon_d = f(\Lambda_{DW}) \tag{2b}$$

The magnetostatic energy arising from the two-dimensionally modulated charges at the upper and lower film surface will be evaluated below with the twin boundary period $\Lambda_{TB}$.

$$\varepsilon_{90°} = \#_{DW} \cdot A_{DW} \cdot \gamma_{90°} = \frac{2\sqrt{2}}{\Lambda_{TB}} \cdot d \cdot \gamma_{90°} \tag{2c}$$

Due to the same amount of 180°-DWs as in the homogeneous band domain pattern, case (a), $\varepsilon_{180°}$ is unchanged. The missing rotation of magnetization leads to the absence of an anisotropy term ($\varepsilon_{ani} = 0$), however on the cost of additional 90°-DWs at each twin boundary. Despite of the *a-c*-twin boundaries no volume charges appear, as the magnetization passes each twin boundary with constant perpendicular component, thus $div(\vec{m}) = 0$. The magnetostatic energy density of the simple band domain pattern, case (a), is now replaced by the energy of a periodic, two-dimensionally modulated charged surface as sketched in Figure 3b with surface charges on the rectangular unit area $\Lambda_{DW} \times \Lambda_{TB}$ given by:

$$\sigma^*(x,y) = J_s \cdot \begin{cases} 0, & 0 \leq x \leq 0.5\Lambda_{DW} \\ -1, & 0.5\Lambda_{DW} \leq x \leq \Lambda_{DW}, 0 \leq y \leq 0.5\Lambda_{TB} \\ +1, & 0.5\Lambda_{DW} \leq x \leq \Lambda_{DW}, 0.5\Lambda_{TB} \leq x \leq \Lambda_{TB} \end{cases} \tag{3}$$

This term can be analyzed through the standard micromagnetic approach[29], namely the calculation of the magnetostatic potential

$$\Phi_d(\vec{r}) = \frac{J_s}{4\pi\mu_0} \cdot \iint\limits_{\substack{extended \\ surface}} \frac{\sigma(\vec{r}')}{|\vec{r}-\vec{r}'|} dS' \tag{4}$$

of the charged surfaces, and subsequent computation of the energy density

$$\varepsilon_d = \frac{J_s}{\Lambda_{DW} \cdot \Lambda_{TB}} \cdot \iint\limits_{S'=\Lambda_{DW} \times \Lambda_{TB}} \sigma(\vec{r}') \cdot \Phi_d(\vec{r}') dS' \tag{5}$$

In Eq. 4 the surface integration has to be computed for the periodically extended surface charge pattern $\sigma(r)$, in Eq. 5 the integration is sufficient on the unit area $\Lambda_{DW} \times \Lambda_{TB}$.

We developed a micromagnetic code which calculates the magnetostatic energy $\varepsilon_d$ of an arbitrary periodic two-dimensional charges pattern based on Eq. 4 and 5. This was successfully tested for the charged patterns of regular perpendicular band domains (BD) and checkerboard domains (CB) and reproduced the known solutions of $\varepsilon_d^{BD} = \frac{1.705}{4\pi} \cdot K_d \cdot \Lambda_{DW}$ and $\varepsilon_d^{BD} = \frac{1.705}{4\pi} \cdot K_d \cdot \Lambda_{DW}$ (see Ref. 16). Applying the code to analyze the charge pattern of case (b) (Eq. 3) we derive again an equation of the form $\varepsilon_d = \frac{\beta}{4\pi} \cdot K_d \cdot \Lambda_{DW}$, with $\beta \approx 0.9$. The variance in the factor $\beta$ is a consequence of the small additional influence of the twinning periodicity on $\varepsilon_d$. Using $\Lambda_{TB} = \delta \cdot \Lambda_{DW}$ ($\delta \approx 0.1 - 0.5$) as observed in the experimental film series[17], results in $\beta \approx 0.88 - 0.95$.

Also for the modulated band domain pattern $\Lambda_{DW}$ affects only the first two energy terms (Eq. 2a/b), and the optimum domain wall period $\Lambda_{DW}$ is therefore again proportional to $d^{1/2}$. The similarity of $\beta \approx 0.9$ with the factor $\alpha = 0.8525$ in case (a) (Eq. 1b) leads to a quantitatively almost identical $\Lambda_{DW}(d)$ behavior, irrespective of the actual domain configuration. Which of the two domain scenarios, case (a) or (b), is energetically favored, depends on the film thickness $d$ and the twin boundary period $\Lambda_{TB}$.

Case (c) is a staircase-like domain pattern which is sketched in Figure 3c. While in bulk the large martensitic variant period allows containing many 180°-DWs, for thin films we consider only one domain wall per martensitic variant, as a higher number would increase the domain wall energy further. In case (c), 180°-DWs run with equal proportion horizontally and vertically through the film. They correspond in area to two vertical 180°-DWs per $\Lambda_{TB}$ extending through the whole film thickness. As in case (b), the twin boundaries constitute 90°-DWs. Internal charges are absent as the flux is guided through the whole depth of the film, leaving only charges at the upper and lower surfaces. These charges form a one-dimensional periodic pattern of the type (+,0,-,0) (see Figure 3c), which gives rise to a magnetostatic energy term. In total we have the following contributions:

$$\varepsilon_{180°} = \frac{2}{\Lambda_{TB}} \cdot d \cdot \gamma_{180°} \quad (6a)$$

$$\varepsilon_d = \frac{3.25}{4\pi} \cdot K_d \cdot \Lambda_{TB} \quad \text{with} \quad \Lambda_{TB} = 0.5 \cdot \Lambda_{DW} \quad (6b)$$

derived again through Eq. 4 and 5 with the appropriate surface charge pattern, and

$$\varepsilon_{90°} = \#_{DW} \cdot A_{DW} \cdot \gamma_{90°} = \frac{2\sqrt{2}}{\Lambda_{TB}} \cdot d \cdot \gamma_{90°} \qquad (6c)$$

The energy terms are formally identical to those in case (b) and now depend solely on the twin boundary period, which has no degree of freedom, but is a given function of the film thickness[17]: $\Lambda_{TB} = 0.2 \cdot d$.

Figure 4a summarizes the total energy for all three cases as a function of the film thickness. Up to a film thickness of 200 nm the homogeneous band domain pattern of completely coupled domains in neighboring variants, case (a), has the lowest total energy $\varepsilon_{tot}$ of all considered domain models. The roughly linear increase in $\varepsilon_{tot}$ is a result of the magnetocrystalline anisotropy contribution and disfavors this domain pattern for larger thicknesses. Above 200 nm, magnetization prefers to follow the local anisotropy axis of the twin variant, which results in an additional, but moderate energy contribution from 90°-DWs and a two-dimensionally modulated charged surface. The magnetostatic energy of this magnetization pattern compares quantitatively with the homogeneous band domain pattern. Accordingly, the equilibrium domain widths of case (a) and (b) (Figure 4b) are essentially unaltered and describe the experimental period.[17] The slight modulation within the band domains observed only in thick films (Figure 2b) indeed suggests that a crossover from case (a) to case (b) occurs at a certain thickness. For a precise confirmation detailed examinations, e.g. of the film cross section, would be required.

Over the whole film thickness range the total energy of the staircase domain pattern, case (c), is clearly above that one of case (b). This originates from the small twin boundary period which introduces a large density of 180°- and 90°-DWs. The consequence is an increased magnetostatic contribution of the one-dimensionally modulated charged surface (see Figure 3c). An explanation for the latter finding can be found in the separation of the charged bands of opposite polarity. Here, flux closure is less effective and the resulting strong stray fields lead to increased magnetostatic energy. One should note that these results do not disagree with the observed staircase domain pattern in bulk material. In macroscopic samples the individual twin variants reach a size which is large enough to contain several magnetic domains for an effective flux closure.

## 4.4 Variant boundaries of two type X variants rotated by 90° around the substrate normal

Four crystallographic equivalent orientations of type X twinned variants exist (Figure 1, orange TB). Accordingly variant boundaries between them are possible. They occur at much larger length scale compared to the twin boundaries within one twinned laminate. In this section, the most obvious variant boundary between two martensitic type X variants is analyzed.

The AFM micrograph in Figure 5a shows the typical periodical surface topography of a 2 μm thick Ni-Mn-Ga film (sample X) exhibiting type X twinning. Two differently oriented martensitic variants are visible and their traces run perpendicular to each other, along MgO[011] and MgO[0$\bar{1}$1]. As the traces of one twinned variant can be transformed to the other one by a rotation of 90° around the substrate normal, we call this type a 90°-variant boundary. This variant boundary is not a strict boundary, but is disturbed and splits up in different segments.

The MFM micrograph in Figure 5b shows the corresponding magnetic band domain pattern, where magnetic band domains run perpendicular to the twin boundary traces within each martensitic variant. Therefore also the magnetic band domains in both variants are aligned perpendicular to each other. At the 90°-variant boundary no correlation between the magnetic domain structures of both variants could be observed. All information obtained from AFM and MFM micrographs are summarized schematically in Figure 5c.

## 4.5 Magnetic domain mirroring at variant boundaries of two type X twinned variants rotated by 180° around the substrate normal

For type X twinned martensitic variants the orientation of traces on the surface is not sufficient for determining the twin boundary orientation, since there are two different ways to tilt the *a-c*-twin boundary from the substrate normal: plus and minus 45° (Figure 1, orange TB). As both variants are connected by a rotation of 180° around the substrate normal we call the variant boundary a 180°-variant boundary.

But how to identify 180°-variant boundaries in AFM measurements, when traces in both neighboring martensitic variants run parallel? For this we first consider the sketch of cross sections through the Ni-Mn-Ga film. In Figure 6 two kinds of 180°-variant boundaries are sketched, which passes through an out-of-plane *a*-axis. In Figure 6a the twinning results in a topography, where the variant boundary exhibits a maximum along the directions of twin boundaries. Also a minimum is possible depending of the orientation of the (101)-twin boundaries (sketched in Figure 6b). The topography of both types of 180°-variant boundaries can be identified in the AFM micrographs (marked as 1 and 2 in Figure 5a, zooms are shown in d and g). The details of the AFM micrographs reveal a topographic ridge (Figure 5d, bright area, corresponds to Figure 6a) and a topographic valley (Figure 5g, dark area, corresponds to Figure 6b). Both 180°-variant boundaries run along the picture diagonals and disturb the typical surface profile only marginally.

For an unambiguous identification of variant boundaries the corresponding MFM micrographs (Figure 5e/h) can be used. In both regions an abrupt inversion of the magnetic contrast of the band domain pattern is observed. This inversion can be attributed to domain mirroring at twin boundaries. For bulk single crystals the domain mirroring effect at one twin boundary had been already described by *Lai* et al.[13] He showed that a mirrored magnetic domain pattern appears at both corresponding sample surfaces connected by a twin boundary at macroscopic distances, but is inverted in contrast. In the present thin films a similar effect occurs, but involves two twin boundaries, which allows observing the pattern on the same (film) surface. As sketched exemplary in Figure 6a, the magnetization $\vec{m}$ (dotted arrows) follows the magnetic easy *c*-axis, which is mirrored at the first twin boundary from out-of-plane to in-plane direction. At the next twin boundary the magnetization is mirrored back to out-of-plane. The magnetization changes from pointing into the film surface to out of the surface on the other side of the twin boundary trace. At the surface an abrupt inversion of the magnetic contrast within the magnetic band domains can be observed. In addition to the both possibilities depicted in Figure 6 two more variant boundaries parallel to a *c*-axis exist (not shown). For these only a topographic contrast, but no mirroring of magnetic domain pattern is expected.

The origin of such 180°-variant boundaries can be explained by multiple transformation nucleuses during the martensitic transformation, where different oriented martensitic variants growth. To keep the crystallography precise, one has to note that the sketches in Figure 6 are simplifications. A general variant boundary, as described in the introductions, involves a slight tilt and rotation of both variants. Accordingly the crystallographic lattice at a variant boundary is expected to be disturbed, which is symbolized by the blue dotted lines in Figure 6. However, these small angular deviations result just in a slight bending of the easy axis, which apparently does not disturb the domain mirroring effect.

## 4.6 Magnetic domain pattern within type Y twinning

In addition to type X twinning another solution for the orientation of {101}-twin boundaries in thin 14M films is possible: The twin planes in type Y twinned martensite run perpendicular through the film plane and parallel to the MgO-substrate edges (see Figure 1, blue TB). Two equivalent solutions for such twin boundaries are possible. In the following the martensitic and magnetic domain pattern of a predominantly type Y twinned epitaxial film (sample Y) is analyzed.

The AFM micrograph in Figure 7a shows a marquetry surface topography. The type Y twinned martensitic variants run as long bands of different length and width parallel to the MgO-substrate edges and perpendicular to each other. Within the bands hardly any topographic contrast is visible (Figure 7c). This is expected for type Y twin boundaries, as for this orientation the twinning angle connecting the *c*-axis of neighboring variants lies within the film plane. Due to the perpendicularly aligned twin boundaries the magnetic easy *c*-axis is always in-plane and just alternates between both substrate diagonals.

The corresponding MFM micrograph of type Y twinned regions in Figure 7b exhibits very low magnetic contrast. A network of lines with a dark magnetic contrast is obvious, which encloses rhombohedral areas. Based on the in-plane orientation of the magnetic easy axis, it is assumed that within these areas no internal charges exist, and stray fields are only created at the clearly visible domain boundaries. The enlarged MFM micrograph (Figure 7d) offers a more detailed picture and reveals on the right hand side of the image a pattern of regularly spaced vertical lines and horizontal zigzag-lines. The most likely corresponding domain pattern is sketched in Figure 7e and displays a pattern of rhombohedral domains. The weaker contrast lines are identified as 90°-DWs running parallel to the MgO-substrate edges, which allow the magnetization to follow the easy axes in each martensitic variant. The zigzag-like contrast can be attributed to 180°-DWs, which are possible as there are two cases for the magnetization direction to follow the easy axis. The length (*l*) of the rhombohedral pattern exceeds their constant width (*w*) and varies without exhibiting any obvious correlation with crystallographic features. As the magnetization passes each domain boundary with constant perpendicular component, no charges are building up and the contrast must originate from the structure of the domain walls itself. The overall domain pattern resembles the magnetic staircase domain pattern in Ni-Mn-Ga bulk[12]. All in all the staircase pattern is just rotated by 45° with respect to the substrate edges since for the thin film experiments the substrate edge is taken as a reference, whereas domain images of bulk single crystals are commonly oriented with respect to the austenitic unit cell. This similarity is plausible, as for the type Y orientation the magnetic easy axis is always in-plane and magnetic stray field effects are thus negligible. While in bulk the spacing of 180°-DWs (*l*) is commonly much shorter than that of 90°-DWs (*w*), this is different for the present thin film. We attribute this to the substantial shorter twin boundary period in films, which fixes *w*, while a large *l* may minimize the total domain wall energy.

## 4.7 Domain pattern at the variant boundary between type X and Y twinned variants

After the detailed analysis of the correlation between the martensitic and magnetic domain pattern in type X and Y twinned 14M martensitic regions, it is interesting to probe the variant boundaries between both types. As an example AFM and MFM micrographs of a 2 µm thick Ni-Mn-Ga film (sample X) containing both types of variants are shown in Figure 8. In the AFM micrograph (Figure 8a) the type Y variants form two long bands running in both equivalent orientations parallel to the substrate edges. This T-shaped region of type Y is surrounded by large type X areas. The presence of both types in one micrograph clearly reveals the substantially different topography. While type X exhibits a pronounced periodical wavy surface topography, only a marginal height contrast can be observed in type Y.

A similar strong difference in contrast of both types is also visible in the corresponding MFM micrograph (Figure 8b). While the type X twinned martensitic variants exhibit magnetic band domains with a high out-of-plane contrast perpendicular to the traces of twin boundaries at the sample surface, the type Y variants show only very little magnetic contrast. Indeed the magnetic contrast within the type Y region is sufficient to resolve the weak zigzag-shape of 180°-DWs, but no additional contrast, which has been clearly visible in the previous enlarged MFM micrograph (Figure 7d).

This mapped region allows analyzing the consequences of variant boundaries between type X and Y on the domain pattern. While within type X regions the band domain period is mostly unaffected when approaching the variant boundary, within the type Y regions the length *l* between the zigzag-patterns vary. From the MFM micrograph it is obvious that the zigzag-patterns usually coincide with band domain boundaries within the type X regions. This suggests that the variability in *l*, described in section 4.6, is used to compensate some of the horizontal component of the band domains, which are fixed in their period (section 4.3). Since type Y variants exhibit only in-plane magnetization, at their variant boundaries towards type X variants no boundary condition for the out-of-plane magnetization component exists. Accordingly two different configurations of adjacent magnetic band domains can be observed in the MFM micrographs. When comparing magnetic band domains at opposite borders of type Y regions both configurations are observed: bright/bright or dark/dark magnetic band domains on both sides (Figure 8b/c, region 2) or antiparallel magnetized out-of-plane band domains (region 1: bright/dark).

## 5  Conclusion

Epitaxial Ni-Mn-Ga films represent a prototype to probe the correlation between martensitic and magnetic microstructure in hierarchically twinned magnetic shape memory alloys. Due to the finite film thickness one has to distinguish two different twin boundary orientations in 14M martensite. In case of perpendicular twin boundaries (type Y) the easy axis alternates between both in-plane orientations. This results in a degenerated staircase-like domain pattern, where the small twinning period and high 180°-domain wall energy inverts the period ratio of 90°- and 180°-domain walls compared to bulk. In case of inclined twin boundaries (type X) the magnetic easy axis alternates between in- and out-of-plane directions. To minimize stray field energy a magnetic band domain pattern is formed. Twin and domain boundaries are aligned orthogonal as this allows minimizing elastic and magnetic energies independently.

Micromagnetic calculations for type X twinned martensite reveal that in agreement with experiments a staircase pattern is not favorable at any film thickness. At thicknesses below 200 nm a band domain pattern originating from a homogeneous magnetization is expected, while above it is favorable for the magnetization to follow the local magnetic easy axis. Despite this expected crossover the domain period remains practically identical and agrees with experiments.

At much larger length scales compared to mesoscopic twinning one observes variant boundaries between laminates with different twin boundary orientations. While in some cases (90°-variant boundary) no correlation of magnetic and martensitic microstructure was observed, domain mirroring occurs at 180°-variant boundaries. At variant boundaries between type X and Y some correlation of domain period is observed which we attribute to the in-plane magnetization components present in both types.

Our analysis reveals how multiferroics are affected by the reduced dimension of thin films. In addition to the increased relevance of magnetic stray field the reduced twining period must be considered. It does not only result in different domain patterns compared to bulk, but also shrinks the size of martensitic laminates. Hence variant boundaries are observable at a higher density and thus are relevant for thin films.

# 6    Acknowledgements

We acknowledge Sandra Kauffmann-Weiss, Christian Behler and Robert Niemann for helpful discussions. This work was funded by the German research foundation (DFG) via the Priority Program SPP1239.

**Figures**

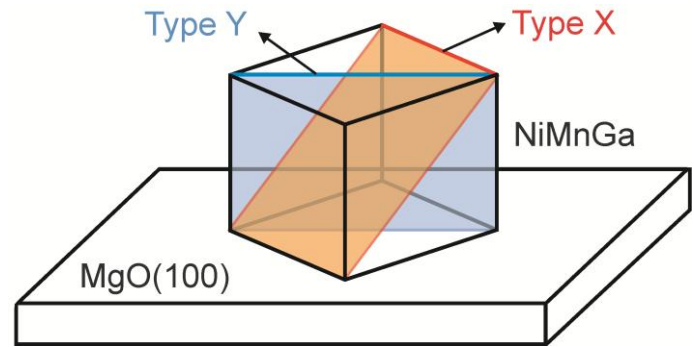

**Figure 1.** *(color online)* Six possible orientations of {110}-twin boundaries (TB) exist, which are sketched here within the austenitic Ni-Mn-Ga unit cell. In thin films two types of twinning have to be distinguished: Type X TBs (orange) run inclined by 45° with respect to the substrate normal and type Y TBs (blue) run perpendicular to the substrate plane.

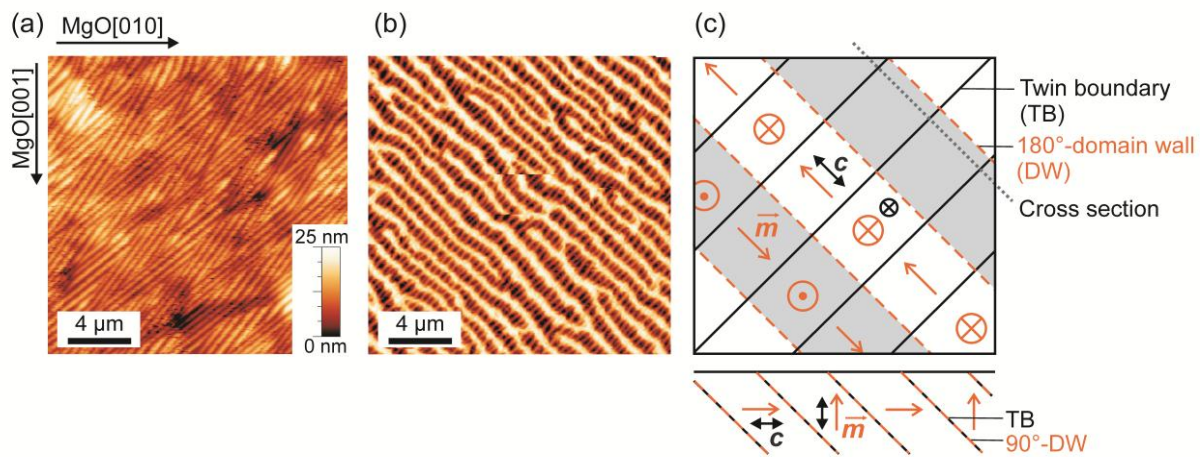

**Figure 2.** *(color online)* Twin and domain pattern of type X orientation. (a) The AFM micrograph of a 2 µm thick Ni-Mn-Ga film (sample X) shows a twinned surface topography where the magnetic easy *c*-axis alternates between in-and out-of-plane orientation. (b) The corresponding MFM micrograph shows magnetic band domains which are aligned perpendicular to the twin boundaries. (c) The crystallographic and magnetic domain structures are sketched as top view and cross section along MgO[0$\bar{1}$1], where the magnetization $\vec{m}$ follows the magnetic easy *c*-axis.[17]

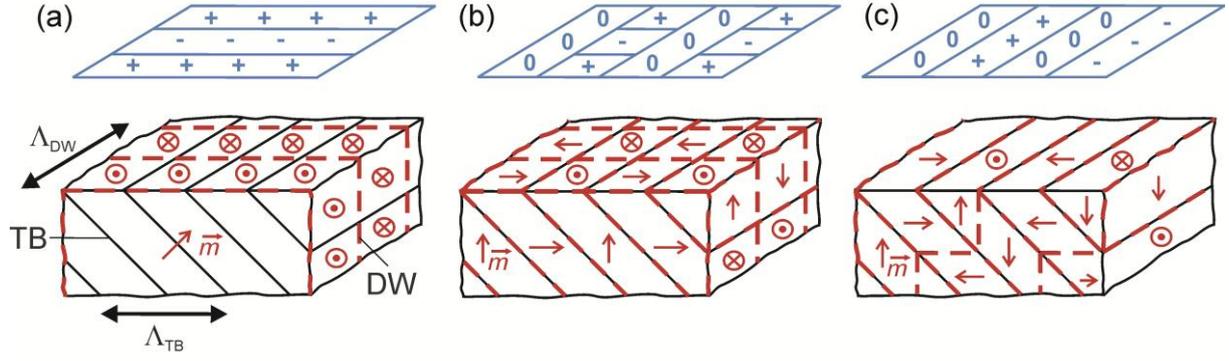

**Figure 3.** *(color online)* Charge distributions and three-dimensional sketches of three possible magnetic domain structures for type X twinning: (a) homogenous band domain model, (b) two-dimensional surface domain model for thin films, (c) staircase-like model for bulk. A micromagnetic comparison of all three cases is described within the text (section 4.3).

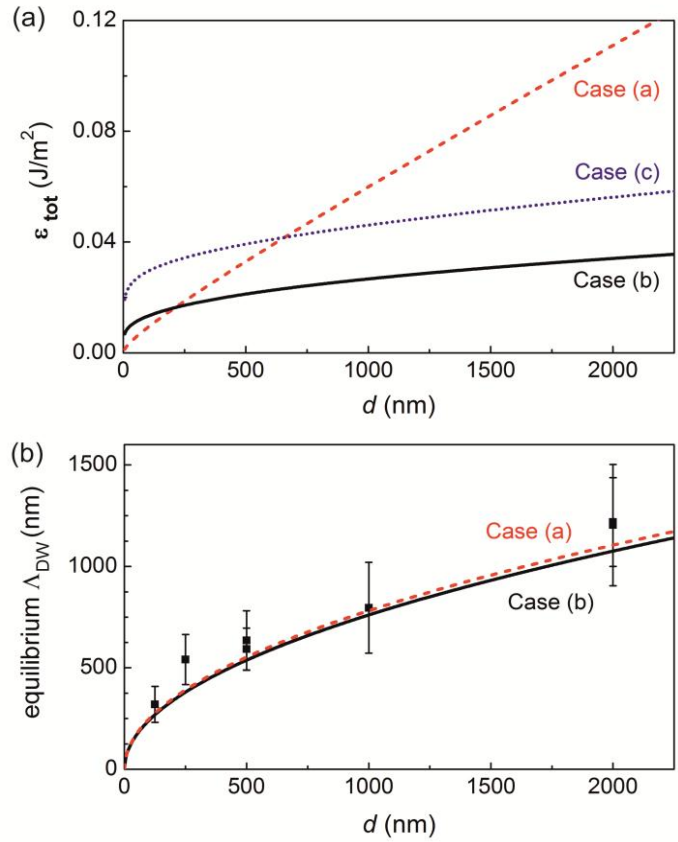

**Figure 4.** *(color online)* (a) Total magnetic energy $\varepsilon_{tot}$ of three different domain models for type X twinning as a function of film thickness $d$: case a) homogeneous band domain pattern (dashed line), case b) two-dimensional surface pattern of thin films (solid line), case c) staircase-like domain pattern (dotted line). (b) The equilibrium domain period $\Lambda_{DW}$ calculated from model case a) (dashed line) and b) (solid line) in comparison with experimental data[17] (squares).

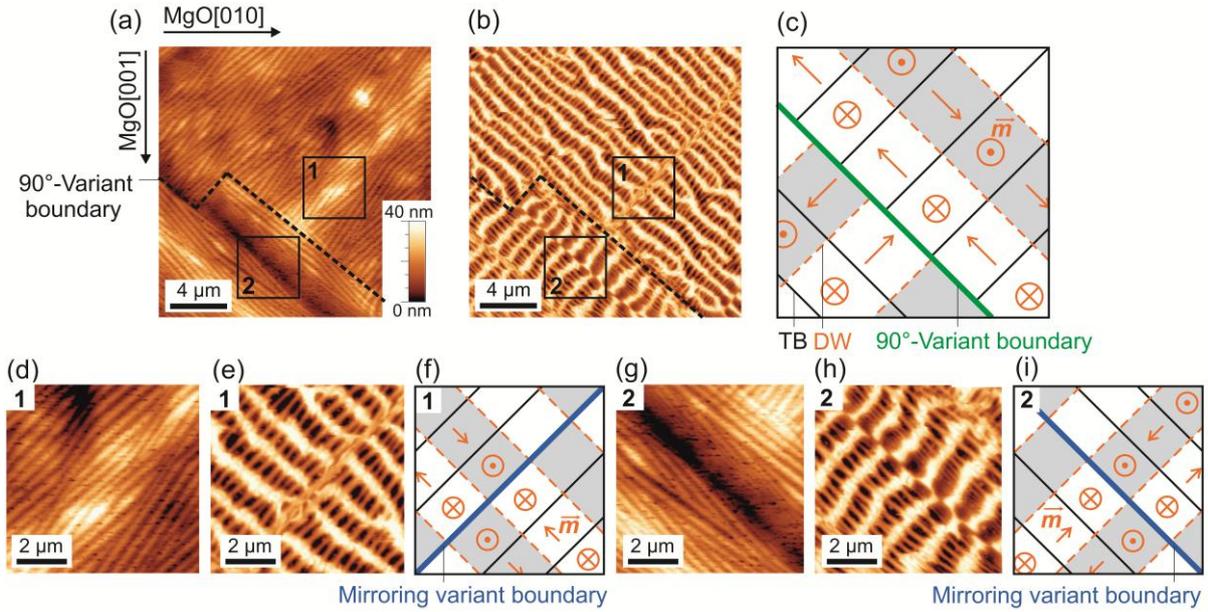

**Figure 5.** *(color online)* Variant boundaries between different orientations of type X twinning. (a) The AFM and (b) MFM micrographs of a 2 μm thick Ni-Mn-Ga film (sample X) show the crystallographic and magnetic domain structure at a variant connecting two type X variants rotated by 90° around the substrate normal. The pattern is summarized schematically in (c). The magnified details reveal a (d) topographic ridge and (g) valley (AFM) whereas in the magnetic contrast (MFM, e/h) an inversion is observed, sketched schematically in (f/i). Both, the topographic features and the domain mirroring, can be explained by a variant boundary connecting two type X variants rotated by 180° around the substrate normal (see Figure 6).

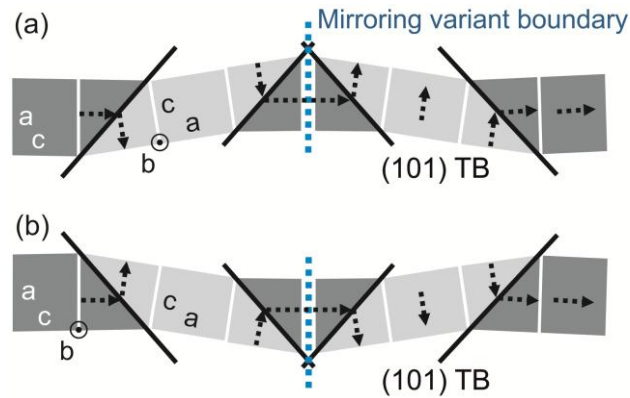

**Figure 6.** *(color online)* Origin of surface topography and domain mirroring at type X variant boundaries. The cross-sections illustrate two types of variant boundaries (blue dotted lines) connecting two type X twinned variants rotated by 180° around the substrate normal. In case that the variant boundary coincides with an *a*-axis (a) a maximum in topography or (b) a valley occurs depending of the twin boundary orientation. The mechanism for domain mirroring is sketched. The dotted arrows illustrate the magnetization direction following the magnetic easy *c*-axis.

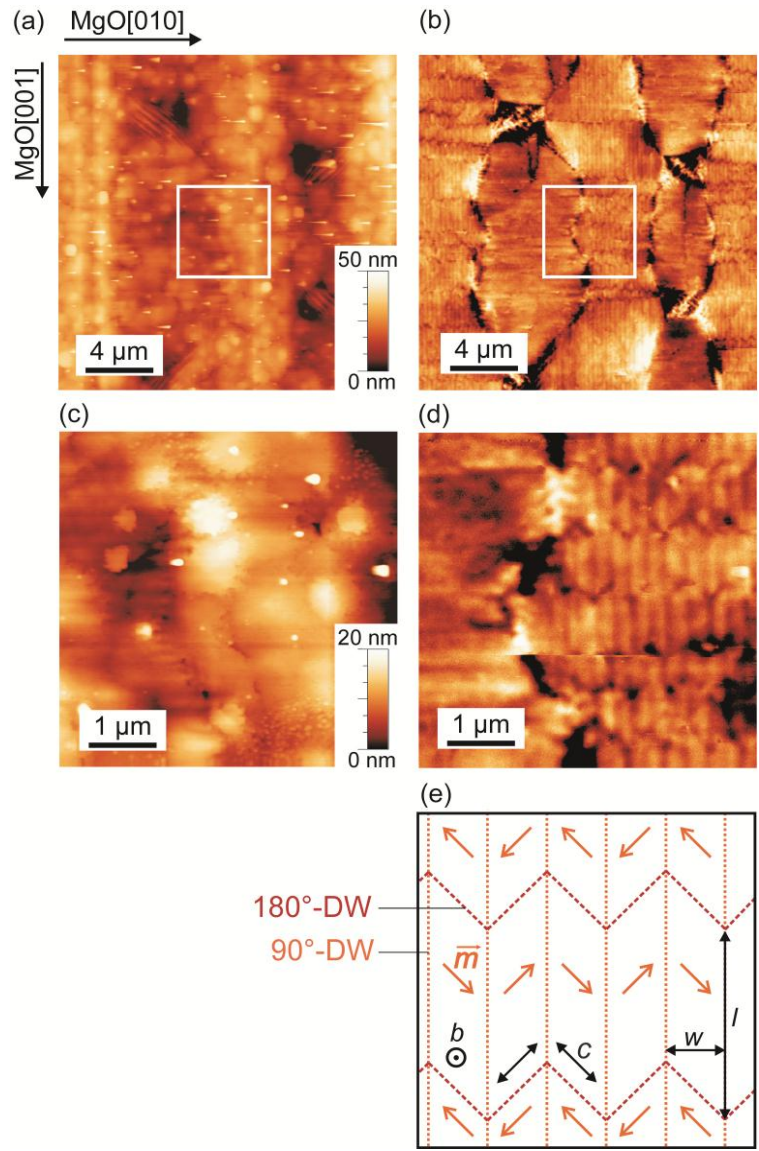

**Figure 7.** *(color online)* The (a) AFM and (b) MFM micrographs of a 1.5 μm thick Ni-Mn-Ga film (sample Y) show the crystallographic and magnetic domain structure of the predominant type Y twinned 14M martensite. Magnified details show a weak (c) topographic (AFM) and (d) magnetic contrast (MFM) within the long variants. (e) Schematic sketch of the magnetic staircase-like domain pattern with 180°- and 90°-domain walls (DW).

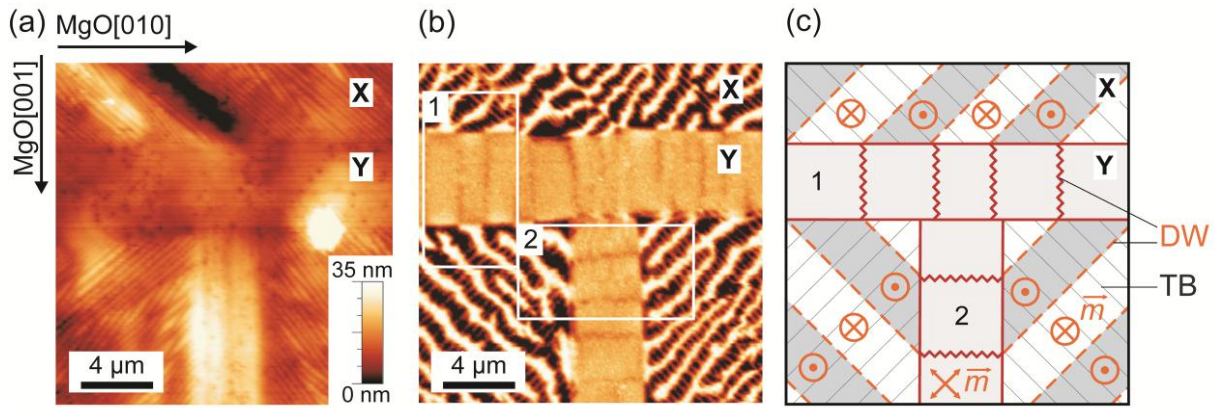

**Figure 8.** *(color online)* Variant boundaries between type X and Y variants. The (a) AFM and (b) MFM micrographs of a 2 μm thick Ni-Mn-Ga film (sample X) show the crystallographic and magnetic domain structure of type X and Y twinned 14M martensitic variants, which is sketched schematically in (c).